\documentclass[12pt]{article}
\usepackage{standard}
\usepackage{psfig}
\newcommand{\pp}{\phantom{-}}
\newcommand{\Li}{\mbox{Li}_2}

\newenvironment{Eqnarray}%
    {\arraycolsep 0.14em\begin{eqnarray}}{\end{eqnarray}}
\newenvironment{Eqnarray*}%
    {\arraycolsep 0.14em\begin{eqnarray*}}{\end{eqnarray*}}

\begin{document}
\thispagestyle{empty}
\begin{flushright}
        MZ-TH/99-27\\
        hep-ph/9911448\\
        November 1999\\
\end{flushright}
\vspace{0.5cm}
\begin{center}
{\Large\bf Inclusive Decays of $ B $ Mesons into $ D_s $ and $ D_s^{*} $}\\[.3cm]
{\Large\bf at $ O(\alpha_s) $ Including $ D_s^{*} $ Polarization Effects}\\[.7cm]
{\large M. Fischer, J.G.~K"orner and M.C.~Mauser}\\[.7cm]
        Institut f"ur Physik, Johannes-Gutenberg-Universit"at\\[.2cm]
        Staudinger Weg 7, D--55099 Mainz, Germany\\[.7cm]
        and\\[0.7cm]
{\large S.~Groote}\\[0.7cm]
        Floyd R. Newman Laboratory of Nuclear Studies\\[.2cm]
        Cornell University\\[.2cm]
        Ithaca, NY 14853, USA
\end{center}
\vspace{0.7cm}


\begin{abstract}\noindent
 The dominant contribution to the inclusive decays of $ B $ mesons into the
 charm-strangeness mesons $ D_s $ and $ D_s^{*} $ is expected to be given by
 the partonic process $ b \!\rightarrow\! c + (D_s^-,D_s^{*-}) $. We determine
 the nonperturbative $ O(1/m_b^2) $ and the $ O(\alpha_s) $ radiative corrections
 to $ b \!\rightarrow\! c + (D_s^-,D_s^{*-}) $ and thereby to the inclusive
 decays $ \bar{B} \rightarrow X_c + (D_s^-,D_s^{*-}) $. The new feature of our
 calculation is that we separately determine the nonperturbative and the 
 $ O(\alpha_s) $ corrections to the longitudinal ($ L $) and transverse 
 ($ T $) pieces of the spin 1 $ D_s^{*-} $ meson. The longitudinal/transverse
 composition of the $ D_s^{*-} $ can be probed through its two principal decay
 modes $ D_s^{*-} \!\rightarrow\! D_s^- + \gamma $ and $ D_s^{*-} \!\rightarrow\!
 D_s^- + \pi^0 $ for which we write down the angular decay distributions.     
\end{abstract}

\newpage


\section{\bf Introduction} 
  
 In a recent paper Aleksan et al. have convincingly argued that the inclusive
 decay $ \bar{B} \!\rightarrow\! X_c + (D_s^-,D_s^{*-}) $ is dominated by the
 partonic process $ b \!\rightarrow\! c + (D_s^-,D_s^{*-}) $ \cite{i1}. The basic
 assumption is that factorization holds for the nonleptonic decay process $
 \bar{B} \!\rightarrow\! X_c + (D_s^-,D_s^{*-}) $. One can then factorize the
 transition into a current-induced $ \bar{B} \!\rightarrow\! X_c $ transition
 and a current-induced vacuum one-meson transition. The leading order contribution
 to the $ \bar{B} \!\rightarrow\! X_c $ transition is given by the partonic
 $ b \!\rightarrow\! c $ transition. Corrections to the leading order result
 set in only at $ O(1/m_b^2) $. They can be estimated using the methods of the
 operator product expansion in HQET.

 Aleksan et al. also pointed out that it would be interesting to experimentally
 measure the longitudinal/transverse composition of the spin 1 meson $ D_s^{*-}
 $ in this inclusive decay which they computed at the Born term level. In an
 accompanying paper the same authors calculated the $ O(\alpha_s) $ corrections
 to the inclusive rates into the spin 0 $ D_s^- $ and the spin 0 $ D_s^{*-}
 $ \cite{i2} without,\footnote{The $ O(\alpha_s) $ corrections to the spin 1
 piece of the weak current keeping both quark masses finite had been calculated
 before in \cite{i3,i4,i5}. The $ O(\alpha_s) $ corrections to the spin 0
 piece of the weak current can be deduced from the corresponding calculation for
 $ t \rightarrow b + H^+ $ \cite{i6}. Latter result had also been used in a
 calculation of the $ O(\alpha_s) $ radiative corrections to $ b \rightarrow c +
 \tau^{-} + \bar{\nu}_{\tau} $ \cite{i7} where the spin 0 piece enters because
 the $ \tau $-mass cannot be neglected in this process.} however, separating the
 longitudinal ($ L $) and transverse ($ T $) contributions in the spin 1 case.
 It is the purpose of this paper to fill the gap left by \cite{i2} and to provide
 analytical formulae as well as the relevant numerical results for the $ O(
 \alpha_s) $ $ L / T $ content of the $ D_s^{*-} $ in this reaction. We emphasize
 that the $ O(\alpha_s) $ corrections calculated here and in \cite{i13} are only
 partial. There are also non-factorizing $ O(\alpha_s) $ corrections as Beneke
 et al. \cite{i14} and Chay \cite{i13} have explicitly shown for the exclusive
 decays $ \bar{B} \!\rightarrow\! \pi \pi $ and $ \bar{B} \!\rightarrow\! 
 D^{(\ast)} \pi^{-} $, respectively. The non-factorizing $ O(\alpha_s) $
 corrections are colour suppressed and are thus expected to be small as e.~g.~
 explicitly shown for $ \bar{B} \!\rightarrow\! D^{(\ast)} \pi^{-} $ in
 \cite{i13}. As concerns the nonperturbative effects we also write down the
 $ L / T $ composition of the nonperturbative $ O(1/m_b^2) $ contribution to the
 $ D_s^{\ast} $ rate as well as to the spin $ 0 $ $ D_s $ rate using results
 of \cite{i12}.
 
 At the Born term level Aleksan et al. found $ \Gamma_L / \Gamma_T \!=\! 1.823 $
 (using their mass values $ m_b \!=\! 4.85 $ GeV and $ m_c \!=\! 1.45 $ GeV). It
 would be interesting to see how radiative corrections and the nonperturbative
 contributions affect this ratio. In the corresponding case $ t \!\rightarrow\!
 b + W^{+} $ (with $ m_b \!=\! 0 $) we found earlier that the ratio $ \Gamma_L /
 \Gamma_T $ is shifted downward by the $ O(\alpha_s) $ radiative corrections by
 an amount of $ 3.5 \% $ \cite{i8}.


\section{\bf Angular decay distributions}

 The longitudinal and transverse content of the diagonal density matrix of the
 $ D_s^{*-} $ (or its charge conjugate state $ D_s^{*+} $) can be determined by
 analysing the angular decay distribution of its subsequent decay into $ D_s^{*-}
 \!\rightarrow\! D_s^{-} + \gamma $ and $ D_s^{*-} \!\rightarrow\! D_s^{-} + 
 \pi^{0} $. The branching ratios into these two principal channels are given by
 $ (94.2 \pm 2.5) \% $ and $ (5.8 \pm 2.5) \% $ \cite{i9}, respectively.
 In terms of the diagonal density matrix elements $ \rho_{mm} $ ($ m \!=\! 0 
 \mbox{($ L $)}, \pm 1 \mbox{($ T $)} $) of the $ D_s^{*-} $ the polar angle
 distribution is given by
 \begin{equation}
  W(\theta) \propto \sum\limits_{m,m^{\prime}} \rho_{mm} \,
  d^{(1)}_{mm^{\prime}}(\theta) \, d^{(1)}_{mm^{\prime}}(\theta) \, 
  |h_{m^{\prime}}|^2. 
 \end{equation}

 \noindent 
 The $ h_m $ are the decay amplitudes of the decays $ D_s^{*-} \!\rightarrow\!
 D_s^{-} + \gamma $ ($ m \!=\! \pm 1 $) and $ D_s^{*-} \!\rightarrow\! D_s^{-}
 + \pi^{0} $  ($ m \!=\! 0 $) where the $ m $ are the magnetic quantum numbers
 of the $ D_s^{*-} $ in the decay frame. The $ d_{m m^{\prime}}^{(1)}(\theta) $
 are the usual Wigner $ d $-function and $ \theta $ is the polar angle of the $ 
 D_s^{-} $ in the $ D_s^{*-} $ rest frame (measured with regard to the original
 momentum direction of the $ D_s^{*-} $) as shown in Fig.~1. One thus obtains
 the polar angle decay distributions
 \begin{equation}
  \frac{d \Gamma_{\bar{B} \rightarrow X_c + D_s^{*-} 
  ( \rightarrow D_s^{-} + \gamma)}}{d \!\cos \theta} =
  \mbox{BR}(D_s^{*-} \rightarrow D_s^{-} + \gamma) 
  \left( \frac{3}{8} (1 + \cos^2 \!\theta) \Gamma_T + 
  \frac{3}{4} \sin^2 \!\theta \, \Gamma_L \right)
 \end{equation}
 and
 \begin{equation}
  \frac{d \Gamma_{\bar{B} \rightarrow X_c + D_s^{*-}
  ( \rightarrow D_s^{-} + \pi^{0})}}{d \!\cos \theta} =
  \mbox{BR}(D_s^{*-} \rightarrow D_s^{-} + \pi^{0})
  \left( \frac{3}{4} \sin^2 \!\theta \, \Gamma_T + 
  \frac{3}{2} \cos^2 \!\theta \, \Gamma_L \right).
  \hspace{5mm}
 \end{equation}

 Considering the fact that the upcoming $B$-factories will be producing upward
 of 10K $ B \bar{B} $ pairs per day and that the inclusive branching ratio of
 the $ B $'s into $ D_s^{*} $'s is expected to lie around $ O(5 \%) $ it should
 not be too difficult to experimentally determine the angular coefficients of
 the two decay distributions and thereby the $ L $/$ T $ content of the $ D_s^{*} $.
 

\section{\bf Born term rates and $ O(\alpha_s) $ radiative corrections}

 Let us begin by writing down the Born term level results for $ b \!\rightarrow\!
 (D_s^{-},D_s^{*-}) + c $ (see Fig.~2a). We shall closely follow the notation of
 Aleksan et al. \cite{i1} throughout. For easy comparison with the numerical
 results of \cite{i1} we shall also adhere to their numerical parameter values.
 One has
 \begin{Eqnarray}
  \Gamma_S^{(0)} \Ss{(b \rightarrow D_s^{-} + c)} & = &
  \frac{G_F^2}{8 \pi} |V_{bc} V_{cs}^{*}|^2 f_{D_s}^2
  \frac{(m_b^2 \!-\! m_c^2)^2}{m_b^2} 
  \Big( \! 1 \!-\! \frac{m_{D_s}^2 (m_b^2 + m_c^2)}{(m_b^2 - m_c^2)^2} \Big)
  p_{D_s} a_1^2, \label{gamma1} \\
  \Gamma_{L+T}^{(0)} \Ss{(b \rightarrow D_s^{*-} + c)} & = &
  \frac{G_F^2}{8 \pi} |V_{bc} V_{cs}^{*}|^2 f_{D_s^{*}}^2
  \frac{(m_b^2 \!-\! m_c^2)^2}{m_b^2} 
  \Big( \! 1 \!+\! \frac{m_{D_s^{*}}^2 
  (m_b^2 \!+\! m_c^2 \!-\! 2 m_{D_s^{*}}^2)}{(m_b^2 - m_c^2)^2} \Big)
  p_{D_s^{*}} a_1^2, \hspace{4mm} \label{gamma2} \\
  \Gamma_{L}^{(0)} \Ss{(b \rightarrow D_s^{*-} + c)} & = &
  \frac{G_F^2}{4 \pi} |V_{bc} V_{cs}^{*}|^2 f_{D_s^{*}}^2
  \frac{(m_b^2 \!-\! m_c^2)^2}{m_b^2} 
  \Big( \! 1 \!-\! \frac{m_{D_s^{*}}^2
  (m_b^2 + m_c^2)}{(m_b^2 - m_c^2)^2} \Big) 
  p_{D_s^{*}} a_1^2. \label{gamma3}
 \end{Eqnarray}
  
 \noindent
 In Eqs.~(\ref{gamma1}-\ref{gamma3}) $ f_{D_s} $ and $ f_{D^{\ast}_s} $ denote the
 pseudoscalar and vector meson coupling constants defined by $ \langle D_s^{-}
 |A^{\mu}|0 \rangle \! = \! i f_{D_s} p^{\mu}_{D_s} $ and $ \langle D_s^{\ast -}
 |V^{\mu}|0 \rangle \! = \! f_{D_s^{\ast}} m_{D_s^{\ast}} \epsilon^{\ast \mu} $,
 respectively. The Kobayashi-Maskawa matrix element is denoted by $ V_{q_1q_2} $
 and the  $ p_{D_s} $ and $ p_{D_s^{\ast}} $ are the three-momenta of the $ D_s $
 and $ D_s^{\ast} $ in the $ b $ rest system. The parameter $ a_1 $ is related to
 the Wilson coefficients of the renormalized current-current interaction and is
 obtained from a combined fit of several decay modes ($ |a_1| \!=\! 1.00 \pm
 0.06 $) \cite{i1}. Note that the structural similarity of the rate formulae for
 the decay into $ D_s $ and the longitudinal $ D_s^{\ast} $ is an accident of
 the Born term calculation and does not persist e.g. at higher orders of
 $ \alpha_s $.
 
 Using $ f_{D_s} \!=\! 230 $ MeV and  $ f_{D_s^{*}} \!=\! 280 $ MeV as in 
 \cite{i1}, $ \tau_B \!=\! 1.6 $ ps, $ V_{bc} \!=\! 0.04 $, $ V_{cs} \!=\!
 0.974 $ and the central value for $ a_1 $ one arrives at
 \begin{equation}
  \mbox{BR}_{b \rightarrow D_s^{- } + c} \cong 3.2 \% \quad
  \mbox{BR}_{b \rightarrow D_s^{*-} + c} \cong 6.8 \% .
 \end{equation}

 Summing up the $ D_s $ and $ D_s^{*} $ modes one arrives at a branching ratio
 of $ 10 \% $ which is consistent with the measured value $ BR(B \!\rightarrow\!
 D_s^{\pm} X) = (10.0 \pm 2.5) \% $ \cite{i9} if one assumes that the above two
 rates saturate the inclusive rate into $ D_s^{\pm} $.

 Next we turn to the $ O(\alpha_s) $ radiative corrections. As explained in 
 \cite{i2}, the radiative gluon corrections connect only to the $ b $ and $ c $
 legs of the parton decay process $ b \!\rightarrow\! (D_s,D_s^{*}) + c $ because
 of the conservation of colour (see Fig.~2b, 2c and 2d). As remarked on earlier
 the radiative corrections for the spin 1 piece are then identical to the
 radiative corrections calculated in \cite{i3,i4,i5} or in \cite{i10} where the
 process $ t \!\rightarrow\! W^{+} \!+\! b $ was considered keeping $ m_b 
 \!\neq\! 0 $. In \cite{i10} we separately computed the radiative corrections to
 longitudinal ($ L $) and transverse ($ T $)  $ W^{+} $'s in the decay process
 and thus these results can directly be transcribed to the present case.

 The two $ L $ and $ T $ pieces can be projected from the hadron tensor by  use of the
 projection operators\footnote{Here we use the notation ``$ T $'' (``transverse'')
 rather than the notation ``$ U $'' (``unpolarized transverse'') used in \cite{i10}.}
 \cite{i10} (we use $ T \!=\! (L + T) - L $ since $ (L + T) $ is simple compared to
 either $ L $ or $ T $)
 \begin{Eqnarray}
  \mbox{I} \! \mbox{P}_{\mbox{\tiny \sl L+T}}^{\mu \nu} & = & 
  \Big( \!\! - g^{\mu \nu} + \frac{q^{\mu} q^{\nu}}{q^2} \Big), \\
  \mbox{I} \! \mbox{P}_{\mbox{\tiny \sl L \phantom{+T}}}^{\mu \nu} & = &
  \frac{q^2}{m_b^2} \frac{1}{|\vec{q} \, |^2}
  \Big( p_b^{\mu} - \frac{p_b \!\cdot\! q}{q^2} q^{\mu} \Big)
  \Big( p_b^{\nu} - \frac{p_b \!\cdot\! q}{q^2} q^{\nu} \Big).
 \end{Eqnarray}
 The scalar spin 0 piece can be obtained with the projector
 \begin{equation} 
  \hspace{-54mm}
  \mbox{I} \! \mbox{P}_{\mbox{\tiny \sl S}}^{\mu \nu} =
  \Big( \frac{q^{\mu} q^{\nu}}{q^2} \Big).
 \end{equation}

 The four-momentum of either the $ D_s $ or the $ D_s^{*} $ is denoted by $
 q_{\alpha} $. The magnitude of the three-momentum of $ q_{\alpha} $ is given by
 $ |\vec{q}| \!=\! \sqrt{q^2 - q_0^2} $.
 We shall not dwell much on the details of our calculation in this short
 communication but refer to \cite{i10} for technical details. Let it be said that
 we use a gluon mass regulator to regularize the infrared singularities differing
 from Aleksan et al. \cite{i2} who use dimensional regularisation instead. We
 shall present our $ O(\alpha_s) $ results in a form where the respective Born
 terms $ \Gamma_i^{(0)} $ are factored out from the $ O(\alpha_s) $ result.
 Including the Born term and the nonperturbative $ O(1/m_b^2) $ contributions to
 be discussed in Sec.4 we write with $ \hat{\Gamma}_S \!:=\! \Gamma_S/\Gamma_S^
 {(0)} $, $ \hat{\Gamma}_{L+T,L} \!:=\! \Gamma_{L+T,L}/\Gamma_{L+T}^{(0)} $ and
 $ \hat{\Gamma}_S^{(0)} \!=\! \hat{\Gamma}_{L+T}^{(0)} \!=\! 1 $,
 $ \hat{\Gamma}_{L}^{(0)} \!:=\! \Gamma_L^{(0)}/\Gamma_{L+T}^{(0)} $
 \begin{equation} 
  \hat{\Gamma}_i = \hat{\Gamma}_i^{(0)} 
  (1 + C_F \frac{\alpha_s}{\pi} \tilde{\Gamma}_i + K_b \, a_i + G_b \, b_i),
 \end{equation}
 where $ i \!=\! S $, $ L + T $, $ L $. $ K_b $ and $ G_b $ are the expectation
 values of the kinetic energy and the chromomagnetic interaction of the heavy quark
 in the $ B $ meson, respectively.

 To begin we list the reduced $ O(\alpha_s) $ rates $ \hat{\Gamma}_i $. For the
 reduced scalar spin 0 rate $ \tilde{\Gamma}_S $ we obtain


 \begin{Eqnarray} 
  \label{defgammas} \tilde{\Gamma}_S & = & 2 \, \Sigma + 1 - 
  \frac{3}{4} \frac{\lambda^{1/2}}{x^2} \ln (w_1 w_{\mu}) \!+\!
  \lambda^{-1/2} B_S^{-1} \Big\{
  \frac{3}{2} \frac{(1 \!-\! y^2) \lambda^{3/2}}{x^2} \ln (y) \!+\!
  y \, (1 \!-\! y)^2 \times
  \nonumber \\ & & \hspace{-0.5cm}
  (1 \!-\! x \!+\! y) (1 \!+\! x \!+\! y) \ln (w_1 w_{\mu}) \!-\! 
  \frac{1}{4} y^2 \, (1 \!-\! x^2)(1 \!+\! y^2) {\cal R}_{\Ss{(-2,-1)}} \!+\!
  \frac{1}{4} (1 \!+\! y^2) \times
  \\ & & \hspace{-0.5cm} 
  (1 \!-\! x^2 \!+\! 3 \, y^2) {\cal R}_{\Ss{(-1,-1)}} \!-\!
  \frac{3}{4} (1 \!+\! y^2) {\cal R}_{\Ss{(0,-1)}} \!-\!
  \frac{1}{2} (1 \!+\! y^2) \, y^2 {\cal S}_{\Ss{(0,0)}} \!+\!
  \frac{1}{2} (1 \!+\! y^2) {\cal S}_{\Ss{(1,0)}} \Big\}, \nonumber
 \end{Eqnarray}
 where we have defined a Born term-like scalar rate by
 \begin{equation} 
  B_S = (1 \!-\! y^2)^2 \!-\! x^2 (1 \!+\! y^2).
 \end{equation}

 \noindent
 The other variables and functions appearing in Eq.~(\ref{defgammas}) are 
 explained at the end of this section. For the total reduced spin 1 rate we
 obtain


 \begin{Eqnarray} 
  \tilde{\Gamma}_{L+T} & = & 2 \, \Sigma - 
  \frac{1}{4} \frac{\lambda^{1/2}}{x^2} \ln (w_1 w_{\mu}) \!+\!
  \lambda^{-1/2} B_{L+T}^{-1} \Big\{
  \frac{1}{2} \frac{(1 \!-\! y^2) \, \lambda^{3/2}}{x^2} \ln (y) \!+\!
  \frac{1}{2} (\lambda \!+\! 6 \, x^2 \, y) \times
  \nonumber \\ & & \hspace{-1cm}
  (1 \!-\! x \!+\! y)(1 \!+\! x \!+\! y) \ln (w_1 w_{\mu}) \!-\!
  \frac{1}{4} y^2 (1 \!-\! x^2)(1 \!+\! 2 \, x^2 \!+\! y^2) 
  {\cal R}_{\Ss{(-2,-1)}} \!+\! \frac{1}{4} ((1 \!+\! x^2 \!-\!
  \nonumber \\ & & \hspace{-1cm} 2 x^4) \!+\! ( 
  4 \!-\! 3 \, x^2) \, y^2 \!+\! 3 \, y^4) {\cal R}_{\Ss{(-1,-1)}} \!-\!
  \frac{1}{4} (3 \!-\! 2 \, x^2 \!+\! 3 \, y^2) {\cal R}_{\Ss{(0,-1)}} \!-\!
  \frac{1}{2} \, (1 \!+\! 2 \, x^2 \!+\! y^2) \times 
  \nonumber \\ & & \hspace{-1cm}
  y^2 \, {\cal S}_{\Ss{(0,0)}} \!+\! 
  \frac{1}{2} (1 \!+\! 2 \, x^2 \!+\! y^2) {\cal S}_{\Ss{(1,0)}} \Big\},
 \end{Eqnarray}
 with
 \begin{equation} 
  B_{L+T} = (1 \!-\! y^2)^2 \!+\! x^2 (1 \!-\! 2 x^2 \!+\! y^2).
 \end{equation}
 
 \noindent
 Finally, the longitudinal piece of the reduced spin 1 rate is given by


 \begin{Eqnarray} 
  \tilde{\Gamma}_L & = &  2 \, \Sigma - 
  \frac{1}{4} \frac{\lambda^{1/2}}{x^2} \ln (w_1 w_{\mu}) +
  \lambda^{-1/2} B_L^{-1} \Big\{
  \frac{1}{2} \frac{(1 \!-\! y^2) \, \lambda^{3/2}}{x^2} \ln (y) +
  y \, (1 \!-\! y)^2 \times
  \nonumber \\ & & \hspace{-0.5cm}
  (1 \!-\! x \!+\! y)(1 \!+\! x \!+\! y) \ln (w_1 w_{\mu}) -
  \frac{1}{4} (1 \!-\! x^2)^3 (1 \!+\! y^2) \, y^2 \, {\cal R}_{\Ss{(-2,1)}} +
  \frac{1}{4} (1 \!-\! x^2) \times
  \nonumber \\ & & \hspace{-0.5cm}
  ((1 \!-\! x^2)^2 \!+\! (6 \!+\! x^2 \!-\! 3 \, x^4) \, y^2 \!+\!
  (5 \!+\! 3 \, x^2) \, y^4) {\cal R}_{\Ss{(-1,1)}} \!-\!
  \frac{1}{4} ((5 \!-\! 2 \, x^2 \!-\! 7 x^4 \!+\! 4 \, x^6) +
  \nonumber \\ & & \hspace{-0.5cm}
  (12 \!-\! 33 \, x^2 \!+\! x^4) \, y^2 \!+\! (7 \!+\! x^2) \, y^4)  
  {\cal R}_{\Ss{(0,1)}} \!+\! \frac{1}{4} ((7 \!-\! 31 \, x^2 \!+\! 4 \, x^4) 
  \!+\! (10 \!+\! x^2) \, y^2 +
  \nonumber \\ & & \hspace{-0.5cm}
  3 \, y^4) {\cal R}_{\Ss{(1,1)}} - \frac{3}{4} (1 \!+\! y^2) 
  {\cal R}_{\Ss{(2,1)}} - \frac{1}{2} ((1 \!+\! 10 \, x^2 \!-\! 11 \, x^4)
  \!+\! (1 \!+\! x^2)^2 y^2) \, y^2 {\cal S}_{\Ss{(0,2)}} +
  \nonumber \\ & & \hspace{-0.5cm}
  \frac{1}{2} ((1 \!+\! 10 \, x^2 \!-\! 11 \, x^4) \!+\!
  (3 \!-\! 4 \, x^2 \!+\! x^4) \, y^2 \!+\! 2 \, (1 \!+\! x^2) \, y^4)
  {\cal S}_{\Ss{(1,2)}} \!-\! \frac{1}{2} ((2 \!-\! 6 \, x^2) +
  \nonumber \\ & & \hspace{-0.5cm}
  (3 \!+\! 2 \, x^2) \, y^2 \!+\! y^4) {\cal S}_{\Ss{(2,2)}} \!+\! 
  2 \, (1 \!+\! y^2) {\cal S}_{\Ss{(3,2)}} \Big\}, 
 \end{Eqnarray}
 where
 \begin{equation} 
  B_L = B_S = (1 \!-\! y^2)^2 \!-\! x^2 (1 \!+\! y^2).
 \end{equation}


 The contribution denoted by $ \Sigma $ is the finite remainder of the Born term
 type one-loop contribution plus the soft gluon contribution. It is given by 
 \begin{Eqnarray} 
  \Sigma & = & \frac{1 \!-\! x^2 \!+\! y^2}{\lambda^{1/2}} 
  \Big\{ \mbox{Li}_2 \Big(1 \!-\! \frac{w_1}{w_{\mu}} \Big) \!-\! 
  \mbox{Li}_2 (1 \!-\! w_1^2) \!-\! \mbox{Li}_2 (1 \!-\! w_1 w_{\mu}) \!+\!
  \frac{1}{8} \ln \Big( \frac{w_{\mu}}{w_1} \Big)
  \ln \Big( \frac{w_{\mu}}{w_1^3} \Big) \!-\!
  \nonumber \\ & & \hspace{-4mm}
  \frac{1}{4} \ln (w_1 w_{\mu}) \Big[
  \ln \Big( \frac{\lambda^{3/2} \, w_1^3 (w_{\mu} \!-\! w_1)}
  {x \, y^2} \Big) \!+\! 1 \Big] \!-\! 
  \ln w_1 \ln \Big( \frac{1 \!-\! w_1^2}{w_{\mu} \!-\! w_1} \Big) \Big\} \!-\! 
  \frac{1 \!-\! y^2}{4 \, x^2} \ln y \!+\!
  \nonumber \\ & & \hspace{-4mm}
  1 \!-\! \frac{1}{2} \ln \Big( \frac{\lambda^2}{x^2 \, y^3} \Big) \!+\! 
  \frac{1}{4} \Big[ \frac{\lambda^{1/2}}{2 \, x^2} \!+\! 
  \frac{x^2 \!-\! 2 y}{\lambda^{1/2}} \Big] \ln (w_1 w_{\mu}) \!-\! 
  \frac{1 \!-\! y^2}{4 \, \lambda^{1/2}}
  \ln \Big( \frac{w_1^3}{w_{\mu}} \Big),
 \end{Eqnarray}
 where we use the abbreviations
 \begin{equation}
  w_{1} := \frac{(1 \!-\! x^2 \!+\! y^2 \!-\! \lambda^{1/2}) \, x}
  {(1 \!+\! x^2 \!-\! y^2 \!+\! \lambda^{1/2}) \, y}, \hspace{2cm}
  w_{\mu} := \frac{(1 \!-\! x^2 \!+\! y^2 \!-\! \lambda^{1/2}) \, x}
  {(1 \!+\! x^2 \!-\! y^2 \!-\! \lambda^{1/2}) \, y},
 \end{equation}

 \noindent
 and $ x \!=\! m_{D_s^{(\ast)}}/m_b $ and $ y \!=\! m_c/m_b $. The kinematical
 factor $ \lambda $ is defined by $ \lambda \!=\! 1 \!+\! x^4 \!+\! y^4 \!-\!
 2 \, x^2 \!-\! 2 \, y^2 \!-\! 2 \, x^2 y^2 $ such that $ p_{D_s^{(*)}} \!=\!
 \frac{1}{2} m_b \, \lambda^{1/2} $.

 The reduced $ O( \alpha_s ) $ rates are given in terms of a set of tree graph
 phase space integrals $ {\cal R}_{\Ss{(m,n)}} $ and $ {\cal S}_ {\Ss{(m,n)}} $
 which are defined by
 \begin{equation} 
  {\cal R}_{\Ss{(m,n)}} := \!\!\! \int\limits_{y^2}^{(1-x)^2} \!\!\!
  \frac{z^m}{\lambda^{n/2}_z} dz, \quad
  {\cal S}_{\Ss{(m,n)}} := \!\!\! \int\limits_{y^2}^{(1-x)^2} \!\!\!
  \frac{z^m}{\lambda^{n/2}_z} \ln \Big(
  \frac{1 \!-\! x^2 \!+\! z \!+\! \lambda_z^{1/2}}
  {1 \!-\! x^2 \!+\! z \!-\! \lambda_z^{1/2}} \Big) \, dz,
 \end{equation}

 \noindent
 where $ \lambda_z \!=\! 1 \!+\! x^4 \!+\! z^2 \!-\! 2 x^2 \!-\! 2 z \!-\! 2 z
 x^2 $. Their solution can be obtained using techniques similiar to the ones
 discussed in \cite{i11}. With the three abbreviations
 \begin{Eqnarray}
  {\cal N}_1 & := & \Li (u \, x) \!-\! \Li \Big( \frac{x}{u} \Big), \\
  {\cal N}_2 & := &  - \ln u \ln (1 \!+\! x) \!+\!
  \ln \Big( \frac{u \!-\! x}{(u \!-\! 1)(1 \!+\! x)} \Big)
  \ln \Big( \frac{u \!-\! x}{u \, (1 \!-\! u \, x)} \Big) \!+\! \\ & - &
  \Li \Big( \frac{1}{u} \Big) \!+\!
  \Li \Big( \frac{u^2 \!-\! 1}{u \, (u \!-\! x)} \Big) \!+\!
  \Li \Big( \frac{1 \!-\! u \, x}{u \!-\! x} \Big), \nonumber \\
  {\cal N}_3 & := & - \ln u \ln (1 \!-\! x) \!-\! 
  \ln \Big( \frac{(u \!+\! 1)(1 \!-\! x)}{u \!-\! x} \Big)
  \ln \Big( \frac{u \!-\! x}{u \, (1 \!-\! u \, x)} \Big) \!+\! \\ & - &
  \Li \Big( \!\!- \frac{1}{u} \Big) \!+\!
  \Li \Big( \frac{(u^2 \!-\! 1) \, x}{u \!-\! x} \Big) \!+\!
  \Li \Big( \!\!- \frac{1 \!-\! u \, x}{u \!-\! x} \Big),
  \hspace{6.2cm}  \nonumber
 \end{Eqnarray}
 one has
 \begin{Eqnarray} 
  {\cal S}_{\Ss{(0,0)}} & = & \lambda^{1/2} \!-\! 2 \, x^2 \ln u \!-\!
  y^2 \ln \Big( \frac{u \!-\! x}{u (1 \!-\! u \, x)} \Big), \\
  {\cal S}_{\Ss{(1,0)}} & = & \frac{1}{4} (1 \!+\! 5 \, x^2 \!+\! y^2)
  \, \lambda^{1/2} \!-\! x^2 (2 \!+\! x^2) \ln u \!-\! \frac{1}{2} y^4 
  \ln \Big( \frac{u \!-\! x}{u (1 \!-\! u \, x)} \Big), \\
  {\cal S}_{(0,2)} & = & - \frac{1}{2 \, x} ({\cal N}_2 \!-\! {\cal N}_3), \\
  {\cal S}_{(1,2)} & = & - \frac{(1 \!+\! x)^2}{2 \, x} {\cal N}_2 \!+\!
  \frac{(1 \!-\! x)^2}{2 \, x} {\cal N}_3 \!+\! {\cal N}_1, \\
  {\cal S}_{(2,2)} & = & - \frac{(1 \!+\! x)^4}{2 \, x} {\cal N}_2 \!+\!
  \frac{(1 \!-\! x)^4}{2 \, x} {\cal N}_3 \!+\! 2(1 \!+\! x^2) {\cal N}_1 \!+\!
  \lambda^{1/2} \!-\! 2 \, x^2 \ln u \!+\! \\ & - & 
  y^2 \ln \Big( \frac{u \!-\! x}{u (1 \!-\! u \, x)} \Big), \nonumber \\
  {\cal S}_{(3,2)} & = & - \frac{(1 \!+\! x)^6}{2 \, x} {\cal N}_2 \!+\!
  \frac{(1 \!-\! x)^6}{2 \, x} {\cal N}_3 \!+\! (3 \!+\! x^2)(1 \!+\! 3 \, x^2)
  {\cal N}_1 \!+\! \frac{1}{4} (9+13x^2+y^2) \lambda^{1/2} \!+\! \\ & - &
  (6 \!+\! 5 \, x^2) \, x^2 \ln u \!+\! \frac{y^2}{2}
  \left( 4(1 \!+\! x^2) \!-\! y^2 \right) 
  \ln \Big( \frac{u \!-\! x}{u (1 \!-\! u \, x)} \Big).
  \nonumber \hspace{4.8cm}
 \end{Eqnarray}

 \noindent The non-logarithmic integrals are given by
 
 \begin{Eqnarray} 
  {\cal R}_{\Ss{(-2,-1)}} & = & \frac{1}{y^2} \, \lambda^{1/2} \!+\! 
  \ln u \!-\! \frac{1 \!+\! x^2}{1 \!-\! x^2} 
  \ln \Big( \frac{u \!-\! x}{1 \!-\! u \, x} \Big), \hspace{7cm} \\
  {\cal R}_{\Ss{(-1,-1)}} & = & \!- \lambda^{1/2} \!-\!
  (1 \!+\! x^2)  \ln u  \!+\!
  (1 \!-\! x^2) \ln \Big( \frac{u \!-\! x}{1 \!-\! u \, x} \Big), \\
  {\cal R}_{\Ss{(\pp 0,-1)}} & = & \frac{1}{2}
  (1 \!+\! x^2 \!-\! y^2) \, \lambda^{1/2} \!-\! 2 \, x^2 \ln u, \\
  {\cal R}_{\Ss{(-2,\pp 1)}} & = &
  \frac{1}{(1 \!-\! x^2)^2} \Big\{
  \frac{\lambda^{1/2}}{y^2} \!+\!
  \frac{1 \!+\! x^2}{1 \!-\! x^2} 
  \ln \Big(\frac{u \!-\! x}{1 \!-\! u \, x} \Big) \Big\}, \\
  {\cal R}_{\Ss{(-1,\pp 1)}} & = & \frac{1}{1 \!-\! x^2} 
  \ln \Big( \frac{u \!-\! x}{1 \!-\! u \, x} \Big), \,\,
  {\cal R}_{\Ss{(0,1)}} = \ln u, \,\,
  {\cal R}_{\Ss{(1,1)}} = \!- \lambda^{1/2} \!+\!
  (1 \!+\! x^2) \ln u, \qquad \\
  {\cal R}_{\Ss{(\pp 2,\pp 1)}} & = & \!- \frac{1}{2} 
  (3 \!+\! 3 \, x^2 \!+\! y^2) \, \lambda^{1/2} \!+\!
  (1 \!+\! 4 \, x^2 \!+\! x^4) \ln u,
 \end{Eqnarray}

 \noindent where $ \Ds{u \!:=\! \frac{1 \!+\! x^2 \!-\! y^2 \!+\! \sqrt{\lambda}}
 {2 \, x}} $.


\section{\bf Nonpertubative contributions}

 When one uses the operator product expansion in HQET one can determine the
 nonpertubative corrections to the leading partonic $ b \!\rightarrow\! c $ rate.
 The nonpertubative corrections set in at $ O(1/m_b^2) $ and arise from the
 kinetic energy and the chromomagnetic interaction of the heavy quark in the
 heavy hadron. The strength of the kinetic and chromomagnetic interactions are
 parametrized by the expectation values of the relevant operators in the
 $ \bar{B} $ system and are denoted by $ K_b $ and $ G_b $, respectively. The
 nonpertubative contributions to the spin 0 and spin 1 rates including the
 $ $ L $/$ T $ $ separation have been calculated in \cite{i12} and can be taken from
 there. One has


 \begin{Eqnarray} 
  S: & & a_{S} = - 1,  \\ & &
  b_{S} = (B_S \lambda)^{-1} \Big[ \!-\! (1 \!-\! y^2)^3 (1 \!-\! 5 y^2) \!+\!
  x^2 (3 \!-\! 7 y^2 \!-\! 11 y^4 \!+\! 15 y^6) \\ & &
  \phantom{b_{S}=(B_S \lambda)^{-1} \Big[} \!-\!
  x^4 (7 \!+\! 10 y^2 \!+\! 15 y^4) \!+\! 5 x^6 (1 \!+\! y^2) \Big], \nonumber \\
  L: & & a_{L} = - 1 - \frac{16}{3} x^2 B_L^{-1}, \\ & &
  b_{L} = (3 B_L \lambda)^{-1} \Big[ \!-\! 3 (1 \!-\! y^2)^3(1 \!-\! 5 y^2) \!-\!
  x^2 (7 \!-\! 27 y^2 \!+\! 65 y^4 \!-\! 45 y^6) \\ & & 
  \phantom{b_{L}=(3 B_L \lambda)^{-1} \Big[} \!+\!
  x^4 (27 \!+\! 34 y^2 \!-\! 45 y^4) \!-\! x^6 (17 \!-\! 15 y^2) \Big], \nonumber \\
  T: & & a_{T} = - 1 + \frac{16}{3} x^2 B_T^{-1}, \\ & &
  b_{T} = 2 x^2 (3 B_T \lambda)^{-1} \Big[ (1 \!-\! y^2)(5 \!-\! 4 \, y^2 \!+\!
  15 y^4) \!+\! x^2 (9 \!+\! 10 y^2 \!+\! 45 y^4) \\ & & 
  \phantom{b_{T}=2 x^2 (3 B_T \lambda)^{-1} \Big[} \!-\! x^4 (29 \!+\! 45 y^2) \!+\!
  15 x^6 \Big], \nonumber \hspace{5mm}
 \end{Eqnarray}
 where
 \begin{equation} 
  B_T = B_{L+T} - B_{L} = 2 x^2 (1 \!-\! x^2 \!+\! y^2).
 \end{equation}
 
 \noindent
 For our numerical evaluations we use $ K_b = 0.013 $ and $ G_b = -0.0065 $ as
 in \cite{i12}.


\section{\bf Numerical results}

 Using $ m_b \!=\! 4.85 \mbox{ GeV} $, $ m_c \!=\! 1.45 \mbox{ GeV} $, $ m_{D_s}
 \!=\! 1968.5 \mbox{ MeV} $ and $ m_{D_s^{*}} \!=\! 2112.4 \mbox{ MeV} $ and
 $ \alpha_s (m_b) \!=\! 0.2 $ we obtain for $ b \!\rightarrow\! c $
 \begin{Eqnarray}
  \hat{\Gamma}_{S \phantom{+L}} & = & \phantom{0.0000 \,}
  ( 1 - 0.09638 - 0.013 \phantom{00} + 0.00467), \\
  \hat{\Gamma}_{L \phantom{+L}} & = &
  0.6459 \, ( 1 - 0.1103 \phantom{0} - 0.03413 + 0.00966), \\
  \hat{\Gamma}_{T \phantom{+L}} & = &
  0.3541 \, ( 1 - 0.1079 \phantom{0} + 0.02553 - 0.02769), \\
  \hat{\Gamma}_{L+T} & = & \phantom{0.0000 \,}
  ( 1 - 0.1095 \phantom{0} - 0.00859 - 0.01803).
 \end{Eqnarray}

 \noindent
 The radiative corrections reduce the rates by about $ 10 \% $, where the
 reduction is rather uniform for the four different rates. The nonperturbative
 corrections range from $ 0.5 \% $ for the chromomagnetic correction to
 $ \hat{\Gamma}_S $ to a maximal $ 3.4 \% $ for the kinetic energy correction
 to $ \hat{\Gamma}_L $ with no uniform pattern in their contributions. At the
 Born term level the transverse/longitudinal composition is given by 
 $ \hat{\Gamma}_T / \hat{\Gamma}_L \!=\! 0.55 $. This ratio is shifted upward
 by the insignificant amount of $ 0.3 \% $ through the radiative corrections.
 Adding all corrections one finds a $ 3 \% $ reduction in the ratio.

 For the $ b \!\rightarrow\! u $ transitions with $ m_u \!=\! 0 $, i.e. 
 $ y \!=\! 0 $ we have
 \begin{Eqnarray}
  \hat{\Gamma}_{S \phantom{+L}} & = & \phantom{0.0000 \,}
  ( 1 - 0.1694 - 0.013 \phantom{00} + 0.00751), \\
  \hat{\Gamma}_{L \phantom{+L}} & = &
  0.7250 \, ( 1 - 0.1777 - 0.02923 + 0.01414 ), \\
  \hat{\Gamma}_{T \phantom{+L}} & = &
  0.2750 \, ( 1 - 0.1150 + 0.02978 - 0.02348 ), \\
  \hat{\Gamma}_{L+T} & = & \phantom{0.0000 \,}
  ( 1 - 0.1605 + 0.00055 - 0.00934).
 \end{Eqnarray}
 
 \noindent
 In the $ b \!\rightarrow\! u $ case the dominance of the longitudinal rate is
 more pronounced. At the Born term level one finds $ \Gamma_T / \Gamma_L \!=\!
 2 x^2 \!=\! 0.38 $. The radiative corrections are no longer as uniform as in
 the $ b \!\rightarrow\! c $ case. Whereas the radiative corrections to 
 $ \hat{\Gamma}_S $, $ \hat{\Gamma}_L $ and $ \hat{\Gamma}_{L+T} $ amount to
 $ 16 \% - 17 \% $, the radiative correction to the transverse rate 
 $ \hat{\Gamma}_T $ is only $ 11.5 \% $. Thus the ratio $ \Gamma_T /
 \Gamma_L $ is shifted upward by $ 7.6 \% $ by the radiative corrections.
 Adding up all corrections one finds a $ 10.4 \% $ upward shift for this ratio.
 Let us mention that our $ O(\alpha_s) $ results on $ \Gamma_{L+T} $ and
 $ \Gamma_S $ numerically agree with the results of \cite{i2} for both the
 $ b \!\rightarrow\! c $ and $ b \!\rightarrow\! u $ transitions.

 As emphasized in the introduction the conclusions drawn in this paper on the
 radiative corrections are tentative in as much as there are also nonfactorizing
 $ O(\alpha_s) $ contributions which have not been included in our analysis.
 Although the nonfactorizing $ O(\alpha_s) $ contributions are colour suppressed
 and thus expected to be small it would nevertheless be worthwhile to try and
 estimate the nonfactorizing $ O(\alpha_s) $ contributions along the lines of
 \cite{i13} and \cite{i14}.

 The last point we want to discuss are the inclusive decays $ \bar{B} 
 \!\rightarrow\! X_C + (\pi^{-},\rho^{-}) $ which can also be induced by the
 diagrams Fig.~2 when the $ c \!\rightarrow\! s $ transition in the upper leg
 is replaced by a $ u \!\rightarrow\! d $ transition. Using $ f_{\pi^{-}} \!=\!
 132 $ MeV, $ f_{\rho^{-}} \!=\! 216 $ MeV and $ V_{ud} \!=\! 0.975 $ one finds
 the Born term branching fractions $ BR_{b \rightarrow \pi^{-} + c} \cong 1.6
 \% $ and $ BR_{b \rightarrow \rho^{-} + c} \cong 4.6 \% $. In the latter case
 the rate is dominated by the longitudinal contribution since $ q^2 \!=\!
 m_{\rho}^2 $ is not far from $ q^2 \!=\! 0 $ where the rate would be entirely
 longitudinal. In fact one finds $ \Gamma_T / \Gamma_L \!=\! 0.067 $. It is
 important to note that the diagrams Fig.~2 are not the only mechanisms that
 contribute to the inclusive decays $ \bar{B} \!\rightarrow\! X_C + (\pi^{-},
 \rho^{-}) $. Additional $ \pi^{-} $ and $ \rho^{-} $ mesons can also be
 produced by fragmentation of the c-quark at the lower leg.\footnote{As
 concerns the inclusive decays $ \bar{B} \!\rightarrow\! X_C + (D_s^{-},
 D_s^{\ast -}) $ the possibility of producing extra $ D_s^{-} $ and
 $ D_s^{\ast -} $ mesons through fragmentation of the $ c $-quark is ruled out
 for kinematic reasons.} As concerns the $ \rho^{-} $ mesons resulting from the
 fragmentation process they would not be polarized along their direction of
 flight. This lack of polarization as compared to the strong polarization of
 the $ \rho $ mesons from the weak vertex could possibly be used to separate
 $ \rho^{-} $ mesons coming from the two respective sources.


 \vspace{1cm}
 {\bf Acknowledgements:} We would like to thank B.~Stugu for an informative
 discussion. M.~Fischer and M.C.~Mauser were supported by the DFG (Germany)
 through the Graduiertenkolleg ``Teilchenphysik bei hohen und mittleren
 Energien'' at the University of Mainz. S.~Groote acknowledges support by the
 Max-Kade Foundation. J.G.~K\"orner was supported in part by the BMBF (Germany)
 through contract 06MZ865.


\newpage
\thispagestyle{empty}
\strut\vspace{6truecm}

\begin{figure}[h]
  \centering \leavevmode 
  \psfig{file=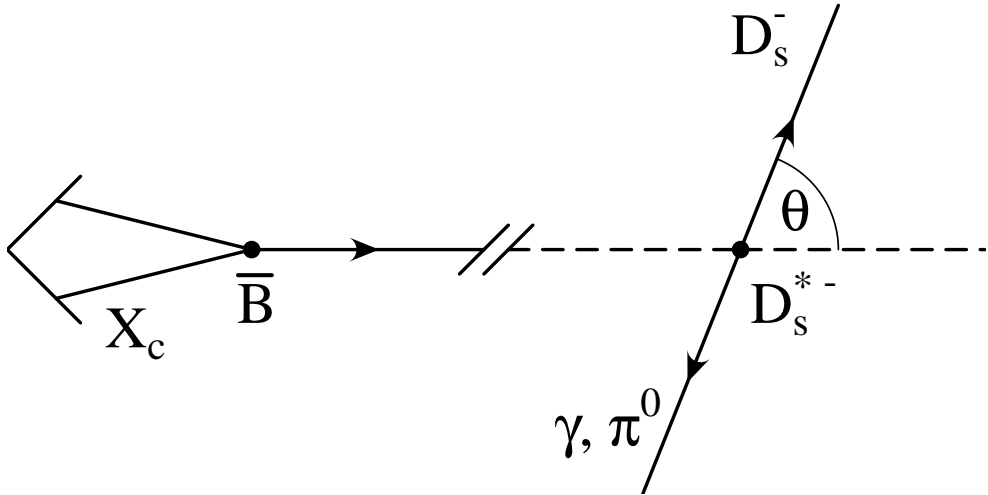,width=13cm,clip=}
  \caption{Definition of polar angle $ \theta $ in the inclusive decay $ \bar{B}
  \!\rightarrow\! X_c + D_s^{*} ( \!\rightarrow\! D_s^{-} + \gamma \mbox{ or }
  \pi^{0} )$. The polar angle $ \theta $ is defined in the $ D_s^{*-} $ rest
  frame relative to the direction of the $ D_s^{*-} $ in the $ \bar{B} $ rest
  frame.}
\end{figure}

\vspace{1truecm}
\centerline{\Large\bf Figure 1}
\newpage
\thispagestyle{empty}
\strut\vspace{3truecm}

\begin{figure}[h]
  \centering \leavevmode
  \put(0,16){a)}
  \psfig{file=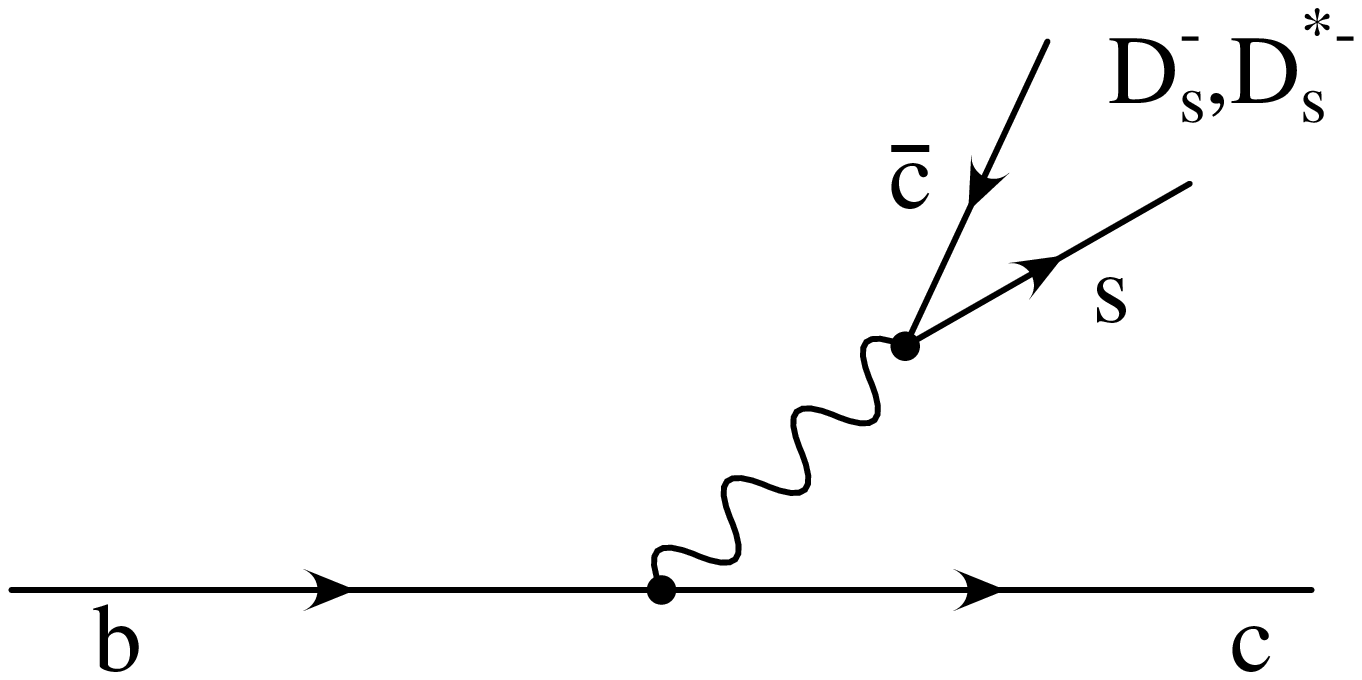,width=7.1cm,clip=}
  \put(0,16){b)}
  \psfig{file=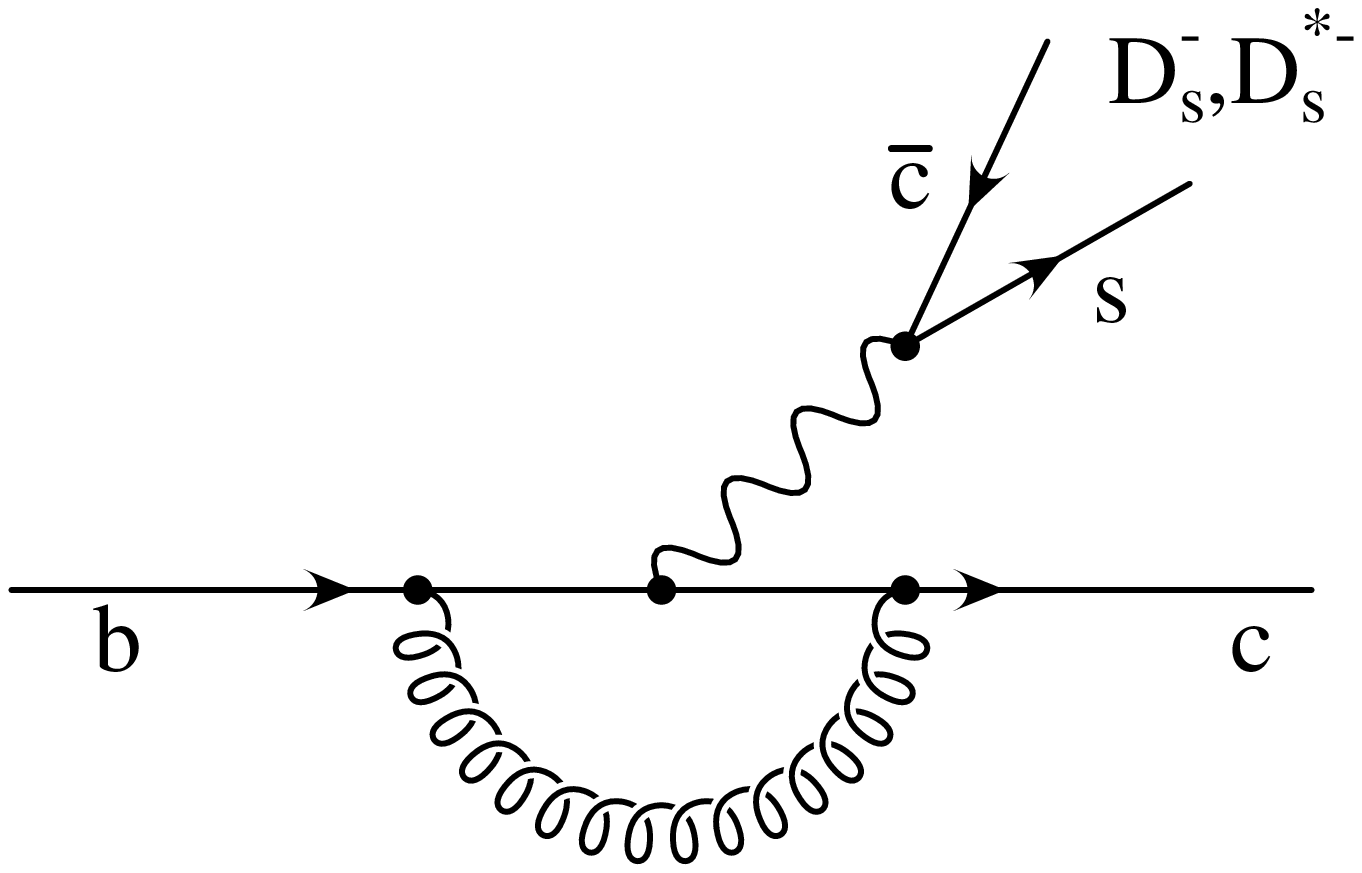,width=7.1cm,clip=}
  \newline
  \put(0,20){c)}
  \psfig{file=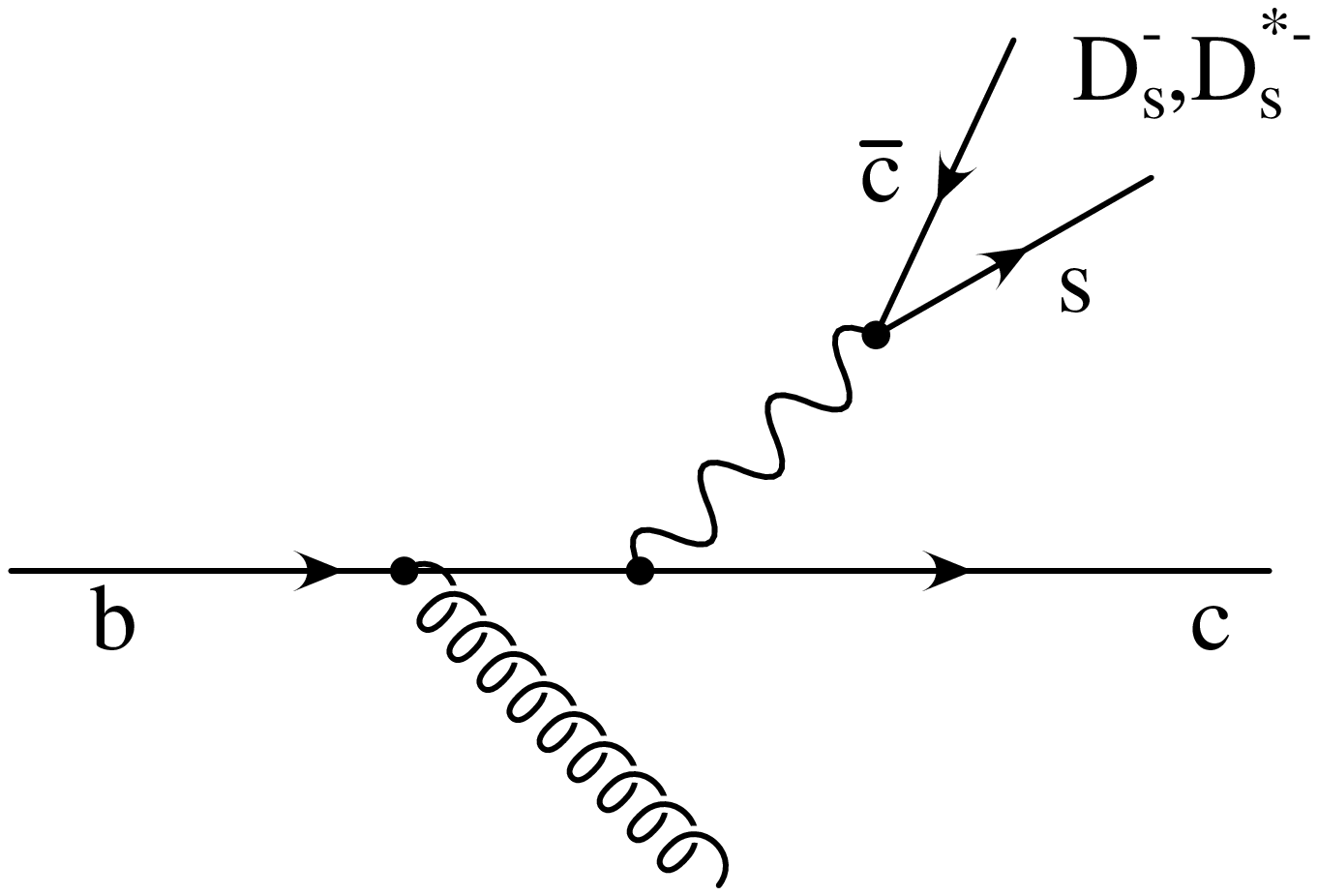,width=7.1cm,clip=}
  \put(0,20){d)}
  \psfig{file=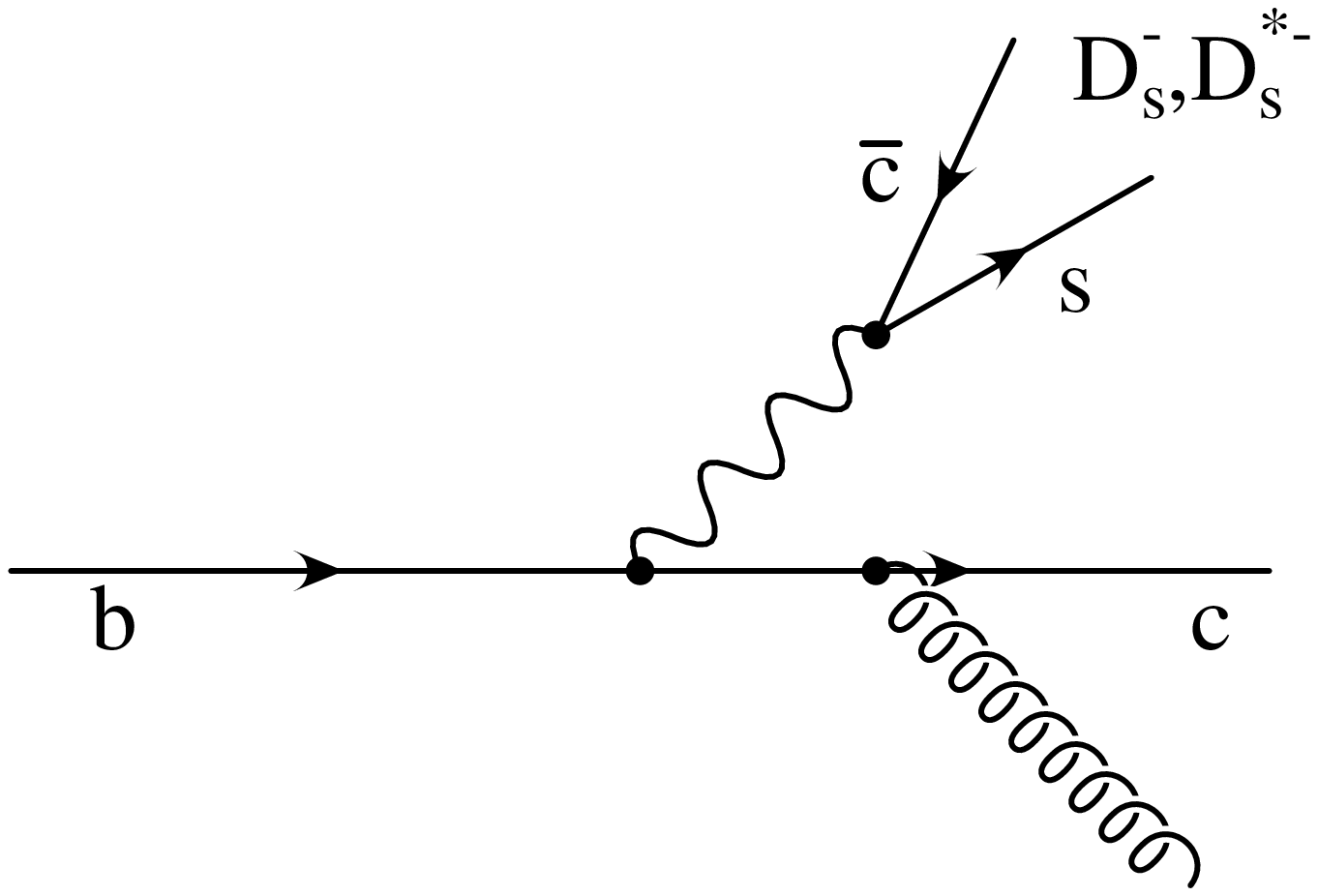,width=7.1cm,clip=}
  \caption{Leading order Born term contribution (a) and $ O( \alpha_s ) $
  contributions (b,c,d) to $ b \!\rightarrow\! c \!+\! (D_s^{-}, D_s^{*-}) $.}
\end{figure}

\vspace{1truecm}
\centerline{\Large\bf Figure 2}


\end{document}